\documentclass[aps,prl,twocolumn,floatfix,nofootinbib,showpacs,superscriptaddress]{revtex4-1}

\usepackage{amsmath,amsfonts,amssymb,bm}
\usepackage{graphicx}
\usepackage{color}
\definecolor{purple}{rgb}{0.5,0,0.5}
\definecolor{blue}{rgb}{0.0,0,0.9}
\usepackage[colorlinks=true, pdfstartview=FitV, linkcolor=purple, citecolor= purple, urlcolor=blue]{hyperref}

\begin{document}

\title{Quark spectral density and a strongly-coupled QGP}

\author{Si-xue Qin}
\affiliation{Department of Physics and State Key Laboratory of
Nuclear Physics and Technology, Peking University, Beijing 100871,
China}

\author{Lei Chang}
\affiliation{Institute of Applied Physics and Computational
Mathematics, Beijing 100094, China}

\author{Yu-xin Liu}
\affiliation{Department of Physics and State Key Laboratory of
Nuclear Physics and Technology, Peking University, Beijing 100871,
China}
\affiliation{Center of Theoretical Nuclear Physics, National
Laboratory of Heavy Ion Accelerator, Lanzhou 730000, China}

\author{Craig D. Roberts}
\affiliation{Department of Physics and State Key Laboratory of
Nuclear Physics and Technology, Peking University, Beijing 100871,
China}
\affiliation{Physics Division, Argonne National Laboratory, Argonne,
Illinois 60439, USA}

\date{17 October 2010}

\begin{abstract}
The maximum entropy method is used to compute the dressed-quark spectral density from the self-consistent numerical solution of a rainbow truncation of QCD's gap equation at temperatures above that for which chiral symmetry is restored.  In addition to the normal and plasmino modes, the spectral function also exhibits an essentially nonperturbative zero mode for temperatures extending to $1.4-1.8$-times the critical temperature, $T_c$.  In the neighbourhood of $T_c$, this long-wavelength mode contains the bulk of the spectral strength and so long as this mode persists, the system may fairly be described as a strongly-coupled state of matter.
\end{abstract}


\maketitle

%
It is believed that a primordial state of matter has been recreated by the relativistic heavy-ion collider (RHIC) \cite{sQGP}.  This substance appears to behave as a nearly-perfect fluid on some domain of temperature, $T$, above that required for its creation, $T_c$ \cite{Song:2008hj}.  An ideal fluid has zero shear-viscosity: $\eta=0$, and hence no resistance to the appearance and growth of transverse velocity gradients.  A perfect fluid with near-zero viscosity is the best achievable approximation to that ideal.  Graphene might provide a room temperature example \cite{graphene}.  From Newton's law for viscous fluid flow; viz., $(v/d) = (1/\eta) (F/A)$, it is apparent that in near-perfect fluids a macroscopic velocity gradient is achieved from a microscopically small pressure.  Strong interactions between particles constituting the fluid are necessary to achieve this outcome.  Hence the primordial state of matter is described as a strongly-coupled quark gluon plasma (sQGP).

Quantum chromodynamics (QCD) produces the bulk of the mass of normal matter.   At $T=0$ it is characterised by confinement and dynamical chiral symmetry breaking (DCSB), phenomena that are represented by a range of order parameters which all vanish in the sQGP.  Understanding the sQGP therefore requires elucidation of the behaviour and properties of quarks and gluons within this state.  Perturbative techniques have been developed for use far above $T_c$; viz., the hard thermal loop (HTL) expansion \cite{Pisarski:1988vd,Braaten:1990wp}, which has enabled the computation of gluon and quark thermal masses $m_T\sim g T$ and damping rates $\gamma_T\sim g^2 T$, with $g=g(T)$ being the strong running coupling.  It also suggests the existence of a collective plasmino or ``abnormal'' branch to the dressed-quark dispersion relation, which is characterised by antiparticle-like evolution at small momenta \cite{Blaizot:2001nr}.

Owing to asymptotic freedom, the running coupling in QCD increases as $T\to T_c^+$.  Therefore, a simple interpretation of the HTL results suggests the plasmino should disappear before $T_c$ is reached because $\gamma_T$ increases more rapidly than $m_T$ and $\gamma_T/m_T\sim 1$ invalidates a quasiparticle picture.
On the other hand, lattice-regularised quenched-QCD suggests that the plasmino branch persists in the vicinity of $T_c$ \cite{Karsch:2009tp}.  It is necessary to resolve the active degrees of freedom in the neighbourhood of $T_c$ because the spectral properties of the dressed-quark propagator are intimately linked with light-quark confinement \cite{Krein:1990sf} and it is the long-range modes which might produce strong correlations.

When addressing issues concerning the dressed-quark propagator it is natural to employ the gap equation, which is one of QCD's Dyson-Schwinger equations (DSEs) \cite{Roberts:1994dr,Roberts:2007jh,Roberts:2007ji}.  Equations of this type are ubiquitous in physics and in QCD the DSEs are distinguished by their ability to unify the analysis of confinement and DCSB within a single nonperturbative, Poincar\'e covariant framework.
Extensive work within this approach has shown that below $T_c$ dressed-gluons and -quarks are confined and chiral symmetry is dynamically broken in the chiral limit \cite{Roberts:2007jh,Roberts:2007ji}, and that deconfinement and chiral symmetry restoration occur via coincident second-order phase transitions \cite{Bender:1996bm,Holl:1998qs,Fischer:2009gk}.  Herein we use the DSEs to elucidate the active fermion quasiparticles for $T\gtrsim T_c$.

On the domain  $T>T_c$ the chiral-limit dressed-quark propagator can be written
\begin{equation}
\label{eq:qdirac}
S(i\omega_n,\vec{p})= -i\vec{\gamma} \cdot \vec{p}\, \sigma_A(\omega_n,
\vec{p}\,^2) - i\gamma_4\omega_n\sigma_C(\omega_n, \vec{p}\,^2) \, ,
\end{equation}
where $\omega_n=(2n + 1){\pi} T$, $n\in \mathbb{Z}$, is the fermion Matsubara frequency.  There is no Dirac-scalar part because chiral symmetry is realised in the Wigner mode.  The retarded real-time propagator is found by analytic continuation
\begin{equation}
\label{eq:qreal}
S^R(\omega,\vec{p})= \left. S(i{\omega_{n}},\vec{p})\right|_{i{\omega_{n}}
\rightarrow\omega+i\eta^+}
\end{equation}
and from this one obtains the spectral density
\begin{equation}
\label{eq:spec}
\rho(\omega, \vec{p})=-2 \Im\,S^R(\omega,\vec{p}) \, .
\end{equation}
Equations~(\ref{eq:qreal}) and (\ref{eq:spec}) are equivalent to the statement:
\begin{equation}
S(i\omega_n,\vec{p}) = \frac{1}{2\pi}\int_{-\infty}^{+\infty}\!\!\!\!\!\!
d\omega^\prime\,\frac{\rho(\omega^\prime,\vec{p})}{\omega^\prime-i{\omega_{n}}
} \, . \label{eq:mat_spec}
\end{equation}
Notably, if one requires a nonnegative spectral density, then Eq.\,(\ref{eq:mat_spec}) is only valid for $T>T_c$; i.e., on the deconfined domain \cite{Bender:1996bm}.

%
%

For an unconfined dressed-quark propagator of the form in Eq.\,(\ref{eq:qdirac}), the spectral density can be expressed
\begin{eqnarray}
\rho(\omega,\vec{p}) = {\rho_{+}} (\omega, \vec{p}\,^2) {P_{+}} +
{\rho_{-}} (\omega, \vec{p}\,^2) {P_{-}}  \,,
\end{eqnarray}
where $P_\pm=(\gamma_{4} \pm i\vec{\gamma}\cdot \vec{u}_p)/2$, $\vec{u}_p \cdot \vec{p} = |\vec{p}|$, are operators which project onto spinors with a positive or negative value for the ratio ${\cal H}:=\,$helicity/chirality: ${\cal H} = 1$ for a free positive-energy fermion.  The spectral density is interesting and expressive because it reveals the manner by which interactions distribute the single-particle spectral strength over momentum modes; and the behaviour at $T\neq 0$ shows how that is altered by a heat bath. As with many useful quantities, however, it is nontrivial to evaluate $\rho(\omega,|\vec{p}|)$.  Nonetheless, if one has at hand a precise numerical determination of the dressed-quark propagator in Eq.\,(\ref{eq:qdirac}), then it is possible to obtain the spectral density via the maximum entropy method (MEM) \cite{Nickel:2006mm}.

We obtain the chiral-limit dressed-quark propagator from the gap equation
\begin{eqnarray}
S(i\omega_n,\vec{p})^{-1} &=& Z_{2}^{A} i\vec{\gamma} \cdot \vec{p}
+Z_{2}i\gamma_4\omega_n+\Sigma^\prime(i \omega_n,\vec{p}) \,, \\
\Sigma^\prime(i \omega_n,\vec{p}) &= & i\vec{\gamma} \cdot \vec{p} \, \Sigma_A^\prime(i \omega_n,\vec{p}) + i\gamma_4\omega_n \,\Sigma_C^\prime(i \omega_n,\vec{p})\,,\\
\Sigma_F^\prime(i \omega_n,\vec{p}) &=&
 T\! \sum_l \int \frac{d^3 q}{(2\pi)^3}
g^2 D_{\mu\nu}({\omega_{n}}\! - \! {\omega_{l}},
\vec{p} \! - \! \vec{q})             \notag    \\
& & \times \frac{1}{3} {\rm tr}_{\rm D}{\cal P}_F\gamma_\mu
S(i\omega_l,\vec{q})\Gamma_\nu(\omega_n,
\omega_l,\vec{p},\vec{q}),\label{eq:gapeq}
\end{eqnarray}
where: ${\cal P}_A = -Z_1^A i\vec\gamma\cdot \vec{p}/\vec{p}\,^2$, ${\cal P}_C = -Z_1 i\gamma_4/\omega_n$; $D_{\mu\nu}$ is the dressed-gluon propagator; $\Gamma_\nu$ is the
dressed-quark-gluon vertex; and $Z_{1,2}$, $Z_{1,2}^A$ are, respectively, the vertex and quark wave function renormalisation constants \cite{Bender:1996bm,Maris:2000ig}.

The gap equation is determined once the kernel is specified.  Herein we work at leading-order in the symmetry-preserving truncation scheme of Ref.\,\cite{Bender:1996bb} and employ a phenomenologically-efficacious one-loop renormalization-group-improved interaction \cite{Maris:2000ig}.  Namely:
\begin{eqnarray}
\nonumber
& & g^{2} D_{\mu\nu}(\omega_{n} - \omega_{l},\vec{p}-\vec{q})
\Gamma_{\nu}({\omega_{n}}, {\omega_{l}},\vec{p},\vec{q}) \\
& = & [{P_{T}^{\mu\nu}}({k_{\Omega}}){D_{T}}({k_{\Omega}})
+{P_{L}^{\mu\nu}}({k_{\Omega}}){D_{L}}({k_{\Omega}})]{\gamma_{\nu}}, \label{eq:model}
\end{eqnarray}
where ${k_{\Omega}}:=(\Omega,\vec{k})=({\omega_{n}} -{\omega_{l}},\vec{p}-\vec{q})$;
\begin{eqnarray}
P_T^{\mu\nu}(k_{\Omega})=\left\{
\begin{aligned}
&0, \qquad \qquad \quad {\mu}\text{ and/or } {\nu} = 4 \, ,  \\
&\delta_{ij}-\frac{{k_{i}} {k_{j}}}{k^2}, \quad {\mu}, {\nu} =1,2,3
\, ,
\end{aligned}
\right.
\end{eqnarray}
with $P_{L}^{\mu\nu} + P_{T}^{\mu\nu} = \delta_{\mu\nu} - k_\Omega^\mu k_\Omega^\nu /k_\Omega^2$; and
\begin{eqnarray}
{D_{T}(k_{\Omega})} &=&\mathcal{D}({k^{2}_{\Omega}},0), \quad
{D_{L}(k_{\Omega})} =\mathcal{D}({k^{2}_{\Omega}},{m_{g}^{2}}) \, ,\\
\nonumber
\mathcal{D}({k^{2}_{\Omega}}, {m_{g}^{2}}) & = & 4{\pi^{2}} D
\frac{s_\Omega}{\omega^6}e^{-s_\Omega/\omega^2}\\
& & + \frac{8{\pi^{2}} {\gamma_{m}}}{{\ln}[ \tau \! + \! (1 \! + \!
s_\Omega/{\Lambda_{\text{QCD}}^{2}} ) ^{2} ] } \,
{\cal F}(s_\Omega)\,,
\end{eqnarray}
with ${\cal F}(s_\Omega) = (1-\exp(-s_\Omega/4 m_t^2)/s_\Omega$, $s_\Omega = \Omega^2 + \vec{k}\,^2 + m^2_g$, $\tau=e^2-1$, $m_t=0.5\,$GeV, $\gamma_m=12/25$, and $\Lambda^{N_f=4}_{\text{QCD}}=0.234$~GeV.  For pseudoscalar and vector mesons with masses$\,\lesssim 1\,$GeV, this interaction provides a uniformly good description of their $T=0$ properties \cite{Maris:2003vk} when $\omega = 0.4\,$GeV, $D=(0.96\,{\rm GeV})^2$.  In generalizing to $T\neq 0$, we have followed perturbation theory and included a Debye mass in the longitudinal part of the gluon propagator: $m_g^2= (16/5) T^2$. 

A justification of the kernel is readily provided.  At $T=0$ it reproduces the results of perturbative QCD for $p^2\gtrsim 2\,$GeV$^2$, so any model-dependence appears only in the infrared, and provides a unified description of light-vector and -pseudoscalar mesons.  It also predicts a momentum dependence for the dressed-quark propagator that is qualitatively in agreement with results from numerical simulations of lattice-QCD \cite{Maris:2002mt}.  The extension to $T>0$ preserves the agreement with perturbative QCD at large spacelike momenta.  Finally, in employing the kernel we obtain coincident second-order deconfinement and chiral symmetry restoring transitions for two massless flavors at $T_c = 0.14\,$GeV,
which is $10$\% smaller than that obtained in Ref.\,\cite{Aoki:2009sc}.

One insufficiency of the interaction defined above is that $D$, the parameter expressing its infrared strength, is assumed to be $T$-independent.  Since the nonperturbative part of the interaction should be screened for $T\gtrsim T_c$, we remedy that by writing $D\to D(T)$ with
\begin{eqnarray}
D(T)=\left\{
\begin{array}{ll}
\displaystyle
D \,, &   T<T_{\text{p}} \, ,  \\
\displaystyle
\frac{a}{b+ \ln[T/\Lambda_{QCD}]}\,, &  T \ge
T_{\text{p}}
\end{array}
\right.\,, \label{DTfunction}
\end{eqnarray}
where $T_{\rm p}$ is a ``persistence'' temperature; i.e., a scale below which nonperturbative effects associated with confinement and dynamical chiral symmetry breaking are not materially influenced by thermal screening.  Logarithmic screening is typical of QCD and with $a=0.028$, $b=0.56$ our numerical solutions yield $m_T = 0.8 \, T$ for $T \gtrsim 2\, T_c$; viz., a thermal quark mass consistent with lattice-QCD \cite{Karsch:2009tp}.  We usually take $T_{\rm p}=T_c$ herein.

\begin{figure}[t]
\centerline{\includegraphics[width=0.47\textwidth]{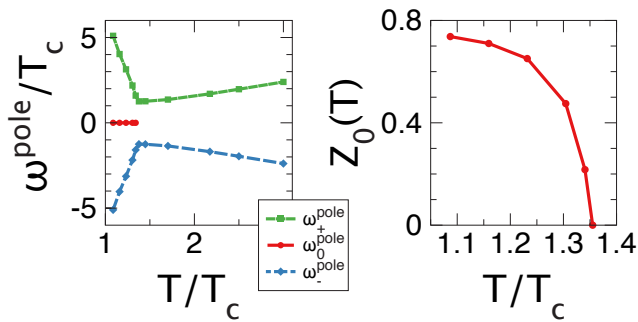}}
%
%
\caption{\label{fig:poles} (Color online)
\emph{Left panel} -- Temperature-dependence of the dressed-quark thermal masses.  Notably, spectral strength is associated with a massless-mode.
\emph{Right panel} -- $T$-dependence of the residue associated with that zero mode.}
\end{figure}

%
We have computed the spectral density by employing the MEM in connection with the solution of our gap equation.  Notably, the behaviour changes qualitatively at $T_c$.  Indeed, employing a straightforward generalisation of the inflexion point criterion introduced in Refs.\,\cite{Roberts:2007ji,Bashir:2008fk}, one can readily determine that reflection positivity is violated for $T<T_c$.  This signals confinement.  On the other hand, the spectral function is nonnegative for $T>T_c$.

In Fig.\,\ref{fig:poles} we depict the $T>T_c$-dependence of the locations of the poles in $\rho(\omega,\vec{p}=0)$; i.e., the thermal masses.  We anticipated that spectral strength would be located at $\omega_+(\vec{p}=0)$ and $\omega_-(\vec{p}=0) = - \omega_+(\vec{p}=0)$, corresponding to the fermion's normal and plasmino modes at nonzero temperature.  However, it is striking that on a measurable $T$-domain, spectral strength is also associated with a quasiparticle excitation described by $\omega_0(\vec{p}=0) = 0$.  The appearance of this zero mode is an essentially nonperturbative effect.  It is an outgrowth of the evolution in-medium of the gap equation's $T=0$ Wigner-mode solution and analogous to this solution's persistence at nonzero current-quark mass in vacuum \cite{Chang:2006bm}.

\begin{figure}[t]
\centerline{\includegraphics[width=0.47\textwidth]{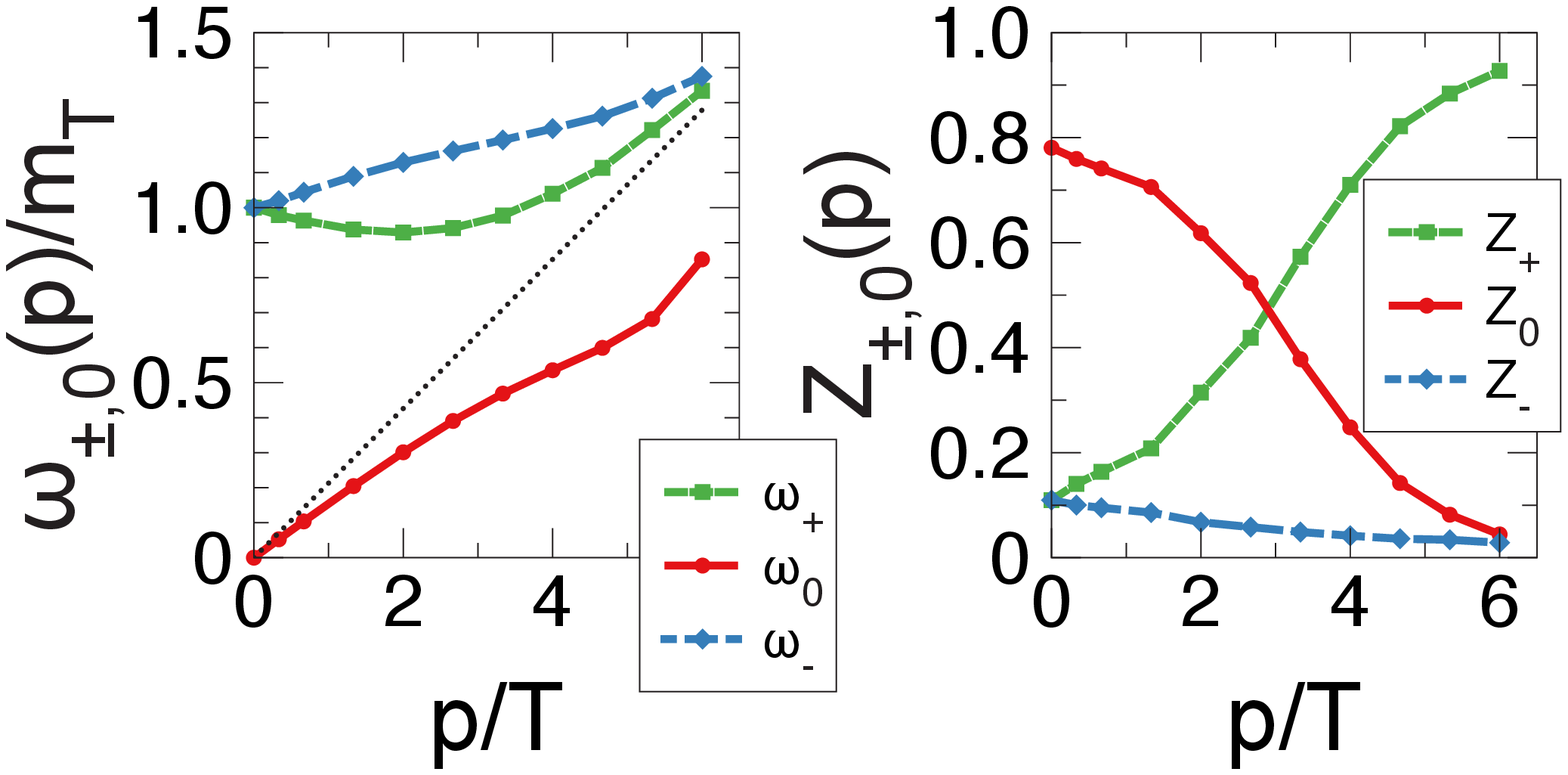}}
\centerline{\includegraphics[width=0.47\textwidth]{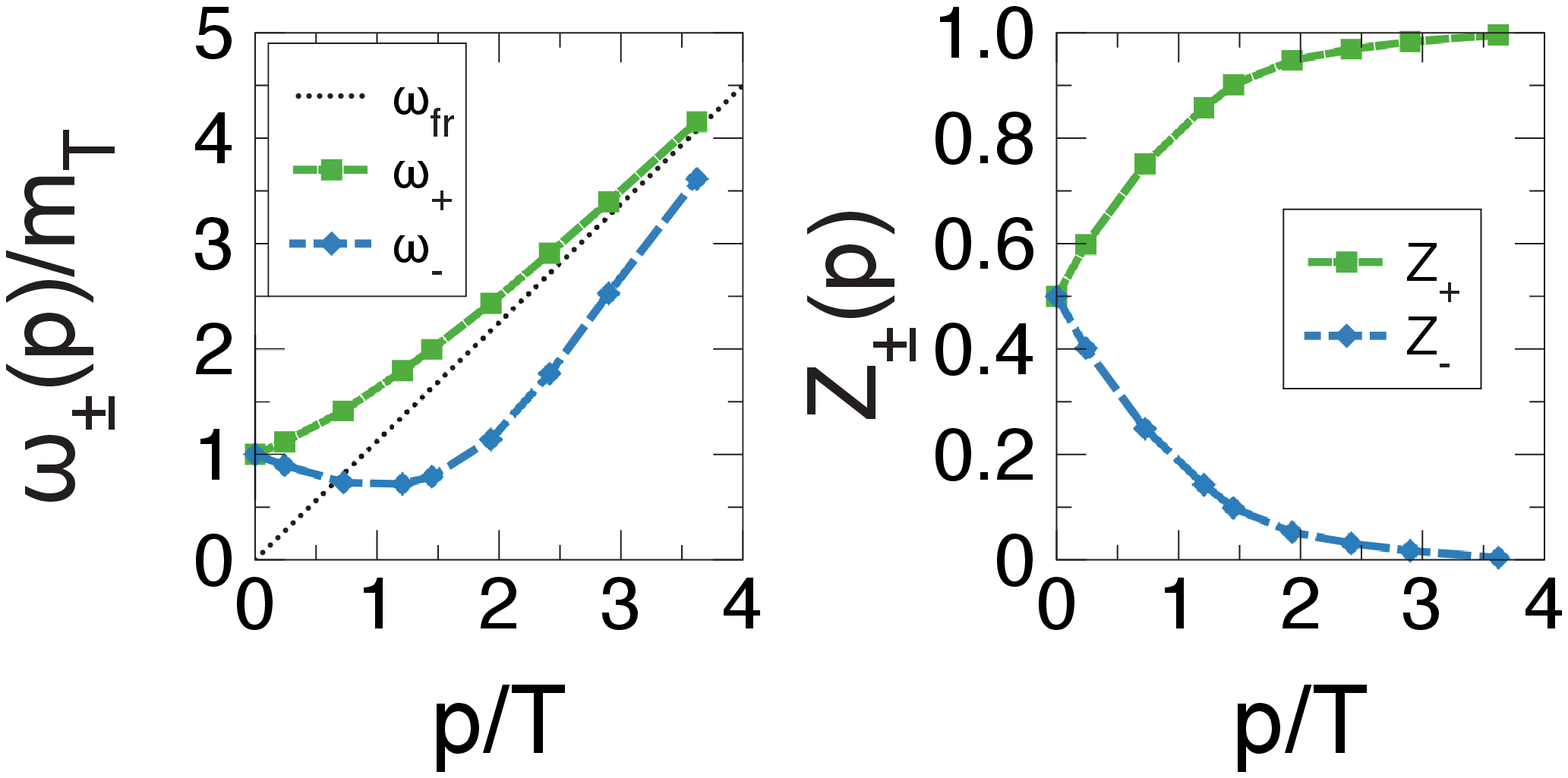}}
\caption{\label{fig:1.1Tc_rel} (Color online)
\emph{Upper panel, left} -- Quasiparticle dispersion relations, $\omega_{\pm,0}(p)$, at temperature $T = 1.1 T_{c}$.  The diagonal dotted line is the free-fermion dispersion relation at this $T$.  \emph{Upper panel, right} -- The residues associated with these quasiparticle poles.
\emph{Lower panel} -- Same information for $T=3 T_c$, whereat the zero mode has vanished.}
\end{figure}

The spectral density possesses support associated with this zero mode on $T\in [0,T_s]$.  In fact: all the Wigner-phase spectral strength is located within this mode at $T=0$; it is the dominant contribution for $T\gtrsim T_c$; and, while it is dominant, it is the system's longest wavelength collective mode.
On the other hand, as evident in the right panel of Fig.\,\ref{fig:poles}, the mode's spectral strength diminishes uniformly with increasing $T$ and finally vanishes at $T_s \approx 1.35\,T_c$.  Then, for $T>T_s$ the quark's normal and plasmino modes exhibit behavior that is broadly consistent with HTL calculations.  This is apparent in Fig.\,\ref{fig:poles} and in a comparison between the upper and lower panels of Fig.\,\ref{fig:1.1Tc_rel}.
Given these observations, we judge that the system should be considered a sQGP for $T\in [T_c,T_s]$, whereupon it contains a long-range collective mode.

We observe that the HTL approach is perturbative and only applicable for $T\gg T_c$.  Hence it could not have predicted the zero mode's existence.  Numerical simulations of lattice-QCD, on the other hand, are nonperturbative.  However, it is practically impossible in contemporary computations to exactly preserve chiral symmetry.  This can plausibly explain the absence of the zero mode in lattice simulations because any source of explicit chiral symmetry breaking heavily suppresses the mode \cite{Chang:2006bm}.

The upper-left panel of Fig.\,\ref{fig:1.1Tc_rel} depicts the dispersion relations for all dressed-quark modes that exist for $T<T_s$ and the behavior of their associated residues.  On this sQGP domain the dispersion relations are atypical, with
\begin{eqnarray}
\omega_{\pm}(|\vec{p}|) &\stackrel{p\sim 0}{=}& m_T
\begin{array}{l}
-0.2\,|\vec{p}|\,,\\
+0.3\,|\vec{p}|\,,
\end{array}\\
%
%
%
\omega_{0}(|\vec{p}|) &\stackrel{p\sim 0}{=}& 0.80 \, |\vec{p}|\,.
\end{eqnarray}
Notwithstanding this, all realise free-particle behaviour for $|\vec{p}|\gg T$.  We note and emphasise that the usual spectral sum rules are satisfied.  Indeed, the identity
\begin{equation}
Z_2^2 \int_{-\infty}^\infty \frac{d\omega^\prime}{2\pi} \; \omega^\prime \rho_\pm(\omega^\prime,|\vec{p}|) = Z_2^A|\vec{p}|
\end{equation}
assists in understanding the momentum-dependence in the upper-left panel of Fig.\,\ref{fig:1.1Tc_rel}.  The upper-right panel displays the momentum-dependence of the pole residues: spectral support is located completely in the normal mode for $|\vec{p}|\gg T$; i.e., on the perturbative domain.

The lower panel of Fig.\,\ref{fig:1.1Tc_rel} characterises the behaviour of $\rho(\omega,|\vec{p}|)$ at on the high-$T$ domain.  In agreement with HTL analysis \cite{Braaten:1990wp}, expected to be valid thereupon, we find only normal and plasmino modes, with
\begin{equation}
\omega_{\pm}(|\vec{p}|) \stackrel{p\sim 0}{=} m_T \pm 0.33 |\vec{p}|\,.
\end{equation}
The plasmino dispersion law exhibits the expected minimum, in this case at $|\vec{p}|/T\simeq 1$; and both $\omega_{\pm}(|\vec{p}|)$ approach free-particle behaviour at $|\vec{p}| \gg T$, with that of the plasmino approaching this limit from below.  The lower-right panel shows that the contribution to the spectral density from the plasmino is strongly damped and contributes little for $p>2 T$.  These results are in-line with those obtained via simulations of lattice-QCD \cite{Karsch:2009tp}.

Equation~(\ref{DTfunction}) is a model and it is natural to enquire after its influence.  None of our results are qualitatively altered by varying $T_{\rm p}$ but, as one would expect, the width of the sQGP domain expands slowly with increasing $T_{\rm p}$; e.g., a 50\% increase in $T_{\rm p}$ produces a 30\% increase in $T_s$.

Whilst we used the MEM to compute $\rho(\omega,|\vec{p}|)$ from the completely self-consistent numerical solution of a rainbow truncation of QCD's gap equation, the appearance of a third and long-wavelength mode in the dressed-fermion spectral density on a material temperature domain above $T_c$ has also been observed in one-loop computations of the fermion self-energy, irrespective of the nature of the boson which dresses the fermion \cite{Kitazawa:2005mp}.  This mode appears for $T>m_G$, where $m_G$ is an infrared mass-scale associated with the boson.  In our case, $m_G = 0.12\,$GeV. Where a comparison is possible, the dependence of our spectral density on $(\omega,|\vec{p}|,T)$ is similar to that seen in the one-loop analyses of model gap equations.  In analogy with a similar effect in high-temperature superconductivity \cite{janko:1997}, that behaviour has been attributed to Landau damping, an interference phenomenon known from plasma physics.

Notably, we find that a coupling to meson-like correlations in the gap equation is not a precondition for the appearance of the zero mode because such correlations are absent in the rainbow truncation \cite{Holl:1998qs}.  On the other hand our gap equation's kernel is characterised by an interaction that features an infrared mass-scale $m_G \lesssim T_c$ and supports dynamical chiral symmetry breaking at $T=0$.  We anticipate that the zero mode will markedly affect colour-singlet vacuum polarisations on $T\in [T_c,T_s]$.  This could be explicated using the methods of Refs.\,\cite{Chang:2008ec}.

Hadron physics experiment has presented us with the fascinating possibility that a near-perfect fluid might have been a key platform in the universe's evolution.  We have sought to provide new insights into this possibility by employing the Dyson-Schwinger equations (DSEs), a tool used widely in many branches of physics and which in QCD provides a nonperturbative, continuum tool that unifies the treatment of confinement, dynamical chiral symmetry breaking and observable phenomena.

In pursuing this aim, we solved self-consistently a rainbow truncation of the gap equation for massless $2$-flavour QCD, employing a kernel whose temperature-dependence is constrained by $T=0$ hadron physics phenomenology and $T\neq 0$ lattice-QCD results, and found that chiral-symmetry restoration and deconfinement occur together at a temperature $T_c= 140\,$MeV.
We subsequently computed the quark spectral density from the gap equation's solution for $T>T_c$ using the maximum entropy method, thereby demonstrating its potential when based on accurate input.
Remarkably, on a significant domain $T/T_c\in [1,1+\Delta]$, $\Delta \simeq 0.4-0.8$, the self-consistently determined Wigner phase supports a zero mode, despite the absence of meson-like correlations in our gap equation's kernel.
This mode contains the bulk of the spectral strength for $T\gtrsim T_c$ and so long as this mode persists, the system may reasonably be described as a strongly-interacting state of matter.
If, as we argued, the existence of this long-wavelength mode is model-independent, then it is natural to anticipate that a strongly-interacting state of matter should precede the QCD phase transition.

Work supported by:
National Natural Science Foundation of China, under contract nos.~10425521, 10705002 and 10935001;
Major State Basic Research Development Program contract no.~G2007CB815000;
and U.\,S.\ Department of Energy, Office of Nuclear Physics, contract no.~DE-AC02-06CH11357.

\end{document}